\documentstyle[prb,aps,multicol]{revtex} 
  \renewcommand{\narrowtext}{\begin{multicols}{2} \global\columnwidth20.5pc} 
  \renewcommand{\widetext}{\end{multicols} \global\columnwidth42.5pc}

\begin{document} 
\title{Experimental tests for the relevance of two-level systems 
for electron dephasing} 
\author{I.~L. Aleiner$^{1,2}$, B.~L. Altshuler$^{3,2}$, and
Y. M. Galperin$^{2,4}$}  
\address{$^{1}$ Department of Physics and Astronomy, SUNY at Stony Brook, 
Stony Brook, NY 11794, USA, \\ 
$^2$Centre for Advanced Studies, Drammensveien 78, 0271 Oslo, Norway,\\
$^{3}$Physics Department, Princeton University, Princeton, NJ 08544,
USA and NEC Research Institute, 
4 Independence Way, Princeton, NJ 08540, USA\\ 
$^4$Department of Physics, University of Oslo, P. O. Box
1048, N-0316 Oslo, Norway and
Division of Solid State Physics, Ioffe
Institute of the Russian Academy of Sciences,
St. Petersburg 194021, Russia  
} 
\date{October 10, 2000}
\maketitle 
 
\begin{abstract} 
The relevance of tunneling two-level systems (TLS)  for electron
dephasing in metals is analyzed. We demonstrate that if the
concentration of TLS  
is sufficient to cause the observed dephasing rate, one also 
should expect quite substantial  effects in the 
specific heat  and ultrasound attenuation.  In both cases TLS 
contribution should dominate the electronic one at low enough 
temperatures.  
\end{abstract} 
\pacs{PACS numbers: }

\narrowtext 
  Physics of two-level systems (TLS) in disordered metals 
has been discussed extensively, see for a review 
Ref.~\onlinecite{Black} and references therein. Recently this field has been 
revisited in connection with 
experiments~\cite{devoret,MW} on  
inelastic relaxation and dephasing of electrons in mesoscopic 
conductors at low-temperatures. Apparent saturation of the 
inelastic relaxation and dephasing rates at $T  \to 0$ is a subject of 
ongoing discussion. This paper is not a review of  the whole spectrum 
of points of view. We analyze only one of the proposed 
explanations,~\cite{imry2,zawadowski1} based on the interaction of 
conducting electrons with defects possessing internal degree of 
freedom. Although the specific structure of the defects can be quite 
complicated, the authors of Refs.~\onlinecite{imry2,zawadowski1} model 
them by TLS. The TLS-induced dephasing has been related to $1/f$ noise
  in Ref.~\onlinecite{imry2}.  
 
In this paper, we examine this explanation 
in comparison with the experimental results on the dephasing. 
In particular, we estimate the TLS concentration, and strength 
of  their coupling with electrons required to describe the 
experimental data. Both the concentration and 
the coupling turn out to be large enough to make noticeable effects on other 
properties of the materials. Namely, we estimate specific heat and 
ultrasonic attenuation in the presence of TLS with the concentration 
that follows from the observed dephasing rate. Observation of these effects 
would be a critical experiment for the TLS-dephasing theory.  
  
Each of the TLS is characterized by a 
diagonal splitting $\Delta$ and tunneling parameter 
$\Lambda$. Introducing Pauli matrices $\sigma_i$ we can write the TLS 
Hamiltonian as  
\begin{equation} 
\tilde{\cal H}_0 = (\Delta/2) \sigma_z + (\Lambda/2) \sigma_x \, . 
\label{TLS-Ham} 
\end{equation} 
Here $\sigma_i$ are the Pauli matrices.
It is convenient to change the 
variables from $\Delta, \, \Lambda$ to energy splitting between the 
two levels, $E=\sqrt{\Delta^2 + \Lambda^2}$, and projection, $p=
(\Lambda/E)^2$. 
As the very existence of TLS 
is caused by disorder, the parameters $E$ and $p$ 
are randomly distributed.  
Assuming that $\Delta$ and $\ln 
\Lambda$ are uniformly distributed and uncorrelated, we obtain 
the conventional distribution of $E$ and $p$ for glassy materials, 
cf. with Ref.~\onlinecite{ahv,phil},   
\begin{equation} 
{\cal P}(E,p)= \frac{P (E)}{2p\sqrt{1-p}} \, . \label{dis} 
\end{equation}  
TLS in glasses are important up to the temperatures $\lesssim 10$~K. 
In the corresponding energy interval the distribution $P (E)$ is 
usually assumed to be a constant. Everything we will be speaking about 
is determined by typical TLS rather than by tails of the distribution. For 
this reason, it is sufficient to  
assume that $P(E)$ is constant in a given energy interval and vanishes 
outside it, $P(E) =(n_T/E^*) \, \Theta(E^* -E)$. Here $n_T$ has a 
meaning of the TLS density. 
In metallic glasses~\cite{Black} $E^* \gtrsim 10-20$ K, and the TLS density of 
states $n_T/E^*$ is determined from the experiments on specific heat 
and acoustic attenuation, $n_T/E^* \approx 10^{16}$ cm$^{-3}$K$^{-1}$. 
 
Let us start our discussion with the 
distribution of these parameters, ${\cal P}(E,p)$, which 
follows from the experimental data on dephasing.~\cite{MW} 
 The striking result of Refs.~\onlinecite{MW} is 
saturation of the dephasing rate $\tau_\varphi^{-1}$ in gold wires as a 
function of temperature below approximately $T_{\max}=1$ K. The authors claim 
that  $\tau_\varphi (T)$-dependence remains weak down at least to  
$T_{\min}=40$ mK. Contribution to $\tau_{\varphi}^{-1}$ of a TLS with 
$T_{\max} \ge E \gtrsim T_{\min}$  is obviously temperature-dependent. 
Consequently, as it was done in Ref.~\onlinecite{imry2}, one has to 
require the energy distribution to be at least three orders of 
magnitude  narrower than that in 
metallic glasses, $E^* \le 
T_{\min}$.  It is not impossible 
that in ordered systems such as 
crystalline wires  TLS are much more similar to each other than in 
glasses and thus have much narrower parameter distributions. However, 
the narrowness of the distribution should manifest itself in physical 
properties.

To describe dephasing by TLS let us include the electron-TLS interaction 
to Eq.~(\ref{TLS-Ham}). A conventional way~\cite{Black} is to express 
it as $\tilde{\cal H}_{\text{int}} =V \sigma_z$ where $V$ has the meaning of 
the difference between scattering potentials in the ``left'' and 
``right'' positions of the defect. After changing the variables to the 
set $(E,p)$ we have  
\begin{equation} 
{\cal H}_{\text{int}} =V\, \sigma_x\, \sqrt{p}+ V\,  \sigma_z\, 
\sqrt{1-p}\,  .  
\label{int-Ham} 
\end{equation} 
As only the first term in Eq.~(\ref{int-Ham}) causes 
dephasing, the contribution of a given TLS is proportional to $p$. As 
a result, the dephasing is determined by TLS with $p \sim 1$. This is 
in contrast with the TLS specific heat where integration over $p$ 
yields a large logarithmic factor.~\cite{ahv,phil}  
 
In contrast with a static defect, TLS causes inelastic scattering of 
electrons, the energy transfer being $E$. The inelastic relaxation 
rate $1/\tau_{\text{in}}$ can be expressed through the 
inelastic scattering cross section, $\sigma_{\text{in}} \equiv 4 \pi 
\chi/k_F^2$ as  
\begin{equation} 
1/\tau_{\text{in}}= v_F \sigma_{\text{in}}n_T=4\pi \chi  v_F n_T/k_F^2\, . 
\label{tf} 
\end{equation}  
Here $v_F$ and $ \hbar k_F$ are the Fermi velocity and momentum, 
respectively. Important dimensionless parameter $\chi \le 1$ has the 
meaning of the inelastic cross section in units of its unitary limit 
$4 \pi/k_F^2$.  
 
 It follows from 
Eq.~(\ref{tf}) that the ratio between the TLS concentration $n_T$ and 
the electron concentration $n_e=k_F^3/3\pi^2$ can be estimated as 
\begin{equation} 
n_T/n_e= (3 \pi/8 \chi)\, (\hbar/\epsilon_F 
\tau_{\text{in}})\, .    
\label{nn} 
\end{equation} 
 
Provided that $E^*\tau_\varphi \gg \hbar$, each inelastic scattering event 
causes dephasing. Therefore, 
the dephasing rate $\tau_{\varphi} \approx \tau_{\text{in}}$. In the 
opposite limit, $E^*\tau_\varphi \ll \hbar$, many inelastic scattering 
events are needed to destroy phase coherence. As a 
result, in general case~\cite{ah,agg}  
\begin{equation} 
\frac{\tau_{\text{in}}}{\tau_\varphi} \sim   
\left\{ 
\matrix{  
1, & E^* \tau_{\text{in}}\gg \hbar \cr  
( E^* \tau_{\text{in}}/\hbar)^{2/3}, &  E^* \tau_{\text{in}} \ll 
\hbar 
}  
\right. \, .  
\end{equation}  
Assuming  $\tau_\varphi=3$ ns and $E^* \approx 20$ mK $> \hbar/(3, 
\text{ns})$, we obtain for $\epsilon_F =5$ eV the estimate $n_T/n_e 
\approx 3 \times 10^{-8}$. For  $n_e \approx 3 \times 10^{22}$ cm$^{-3}$ 
the   TLS density of states turns out to be of the order of 
$(n_T/E^*,\text{cm}^{-3}\text{K}^{-1}) \approx 5 \times 10^{16}/\chi$.
At $E^* \tau_\varphi \ll \hbar$ the  above estimate should be 
multiplied by a large factor $(\hbar/E^* \tau_{\varphi})^2 \gg 1$.  
  
Let us compare the contributions to the specific 
heat of TLS, $C_T$, and electrons, $C_e$: 
\begin{eqnarray} 
&&C_e=3 n_e T/2 \epsilon_F\, ,   
\label{ce} 
\\ 
&&C_T=\frac{n_T T {\cal L}}{E^*} f \! \left({E^* \over T}\right), \ f 
(x)= \left\{  
\matrix{ 
1, & x \ll 1 \cr x^3 / 12, & x \ll 1  
} \right. . 
\label{ct} 
\end{eqnarray} 
Large factor ${\cal L} \gg 1$ appears in Eq.~(\ref{ct}) due to the 
fact that, in contrast with the case of $1/\tau_\varphi$, even very 
``slow'' TLS contribute to the specific heat. We think that this fact 
was not fully appreciated in Refs.~\onlinecite{imry2} and 
\onlinecite{zawadowski1}.   
Formally, $\cal L$ originates from the  
logarithmic divergence of the integral $\int {\cal P}(E,p)\, dp$ at $p 
\to 0$. The 
actual  limits of this integration  
are not well understood. Usually this factor is estimated as ${\cal L} 
= \ln (t_{\text{exp}}/\tau_{\min})$ where $t_{\text{exp}}$ is the 
measurement time while $\tau_{\min}$ in the minimal relaxation time 
of TLS population. A realistic estimate is ${\cal L} \approx 20-40$.

It follows from  Eqs.~(\ref{ce}), (\ref{ct}), and (\ref{nn}) that 
at  high temperatures $C_T$ is negligible, whereas at $T \ll E^*$  
\begin{equation} 
C_T/C_e \approx ({\cal L}/\chi)\, (\hbar/E^* \tau_{\text{in}})\, . 
\label{cc} 
\end{equation} 
For our example, $C_T/C_e \approx 0.1 {\cal L}/\chi $.  
 
The dimensionless crossection $\chi$ can be 
estimated from relaxation acoustic attenuation, see, 
e.g., Ref.~\onlinecite{Black} and discussion below. In metallic glasses 
such an estimation gives $\chi = 0.01 - 0.1$. Taking ${\cal L} =30$, $\chi 
= 0.03$ we conclude that $C_T$ at low enough temperatures exceeds $C_e$ by
\emph{about hundred times}.  At the same time, as $T 
\gg E^*$ the TLS contribution rapidly decreases with temperature increase.   
Consequently, the ratio $C/T$  changes dramatically at $T 
\approx E^*$, being almost constant both at $T \ll E^*$ and $T \gg 
E^*$. This effect provides a possibility to experimentally determine 
both $E^*$ and $n_T$. 

 
Another independent way to determine parameters of TLS is the study of the
 acoustic attenuation. The advantage is that parameters $\chi$ 
and $E^*$ are relevant even at $T \gg E^*$. Moreover, it is 
well known~\cite{Black} that there is almost no ultrasound attenuation 
due to free electrons and phonons at low enough acoustic frequency and 
temperature. Therefore even low concentration of TLS provides 
dominating contribution. 
 
 There are two 
contributions of TLS to the attenuation\cite{Hunklinger}. 
The first one  is a 
direct inter-level absorption of acoustic quanta  known as the 
resonant mechanism. The second, relaxation mechanism arises from 
acoustic wave-induced time dependence of the energy splitting 
$E$. Nonequilibrium component of the TLS population which is caused 
by this dependence relaxes due to electron-TLS interaction. This 
relaxation produces acoustic attenuation.  
 
To evaluate the power $W$ dissipated per unit volume of the metal we 
make a usual assumption\cite{Hunklinger} that the ultrasound 
affects only the diagonal part of the Hamiltonian 
Eq.~(\ref{TLS-Ham}): $\Delta \sigma_z$ is substituted by 
$(\Delta +a \cos \omega t)\sigma_z$ where $\omega$ is the ultrasound 
frequency. In the basis where the unperturbed Hamiltonian, 
Eq.~(\ref{TLS-Ham}), is diagonal,   the first-order perturbation can be 
written as a sum  
$a\, \cos \omega t\, \left(\sigma_z\sqrt{1-p}+ \sigma_x\sqrt{p} \right)$.   
The two terms 
lead, respectively, to the relaxation and resonant attenuation, 
$W_{\text{rel}}$ and $W_{\text{res}}$. 
Note that the off-diagonal part of the perturbation vanishes  
at $p \to 0$, and thus the integral over $p$ is determined by $p \sim 
1$. Therefore $W_{\text{res}}$ in contrast with $C_T$ does not contain 
the large logarithmic factor $\cal L$. This factor does not appear in 
$W_{\text{rel}}$ either for the same reason as it did not appear in 
$1/\tau_\varphi$ -- TLS-electron relaxation rates vanish at $p \to 0$.

In the absence of  TLS relaxation one can calculate the contribution
to $W$  from one impurity $i$ characterized by $(p_i,E_i)$ 
using the Fermi golden rule  
\begin{equation} 
\delta W_{\text{res}}^{(i)}(\omega)=(\pi a^2/2)\,\omega p_i\tanh (E_i /2T) 
\, \delta\left( \hbar 
\omega -E_i 
\right) \, .
\label{Q1} 
\end{equation} 
Here  $\hbar\omega$ is the energy transfer, and 
the factor $p_i$ appears because the
transition matrix element is proportional to the tunneling coupling
$\Lambda$, while the factor $\tanh (E_i /2T)$ is the occupation
difference of the two levels with the distance $E_i=\hbar \omega$.
 
Interaction of TLS with electrons broadens the resonance.  One can 
take this broadening into account replacing the $\delta$-function in 
Eq.~(\ref{Q1}) by Lorentzian $\pi^{-1}\Gamma^{(i)}/[(\hbar \omega 
-E^{(i)})^2 + (\Gamma^{(i)})^2]$. Here $\Gamma$ is the TLS off-diagonal 
relaxation rate which is an analog of the rate $\hbar/T_2$ in the 
physics of spin resonances. As each inelastic process leads to 
simultaneous relaxation of a TLS and an electron, the rate  $\Gamma$
is directly  
related to the electron inelastic relaxation rate, Eq.~(\ref{tf}), and 
thus is determined by the same dimensionless inelastic crossection 
$\chi$, see Ref.~\onlinecite{Black}, 
\begin{equation} 
\Gamma^{(i)} =\chi\, p_i E_i/[2\tanh(E_i/2T)]\, . 
\label{Gamma} 
\end{equation}  
 
The relaxation contribution is given by the Debye-type formula, see 
e.~g. Ref.~\onlinecite{Jackle}, 
\begin{equation} 
\delta W^{(i)}_{rel} 
= \frac{\omega a^2}{2}\cdot \frac{\omega\, (1-p)}{4T\cosh^2(E_i/2T)}\cdot 
\frac{2 \Gamma^{(i)}/\hbar} 
{\omega^2+ (2 \Gamma^{(i)}/ \hbar)^2}. 
\label{Q2} 
\end{equation} 

The total dissipated power per unit volume 
is obtained by averaging  
of Eqs.~(\ref{Q1}) and~(\ref{Q2}) 
over $p_i$ and $E_i$ with the distribution (\ref{dis}). 
 For presenting the results, 
we introduce dimensionless frequency, $\Omega$, and 
temperature, $\theta$, 
\begin{equation} 
\Omega = \hbar \omega/E^*\, , \quad \theta = 2T/E^*\, . 
\label{not} 
\end{equation} 
In these terms the total dissipated power per unit volume $W$ has the form 
\begin{equation} 
W= (n_Ta^2/2\hbar)\, \Omega \,F (\chi,\Omega,\theta)\, , \ 
F=F_{\text{res}}+F_{\text{rel}}\,. \label{Qtot} 
\end{equation} 
where the dimensionless functions are given by 
\begin{eqnarray} 
F_{\text{res}}= 
\int_0^1\int_0^1 \frac{dxdp\,  \chi xp\, (1-p)^{-1/2}}{4(x-\Omega)^2 + (\chi xp)^2 \tanh^{-2}(x/\theta)} \, .  
\label{fres}\\ 
F_{\text{rel}}= \int_0^{1/\theta}\frac{ \chi \theta y\, dy\, 
}{2 \cosh^3 y}   \int_0^1  \, \frac{ dp \, \sqrt{1-p}\, \Omega\, \sinh y}{\Omega^2 
\tanh^2 y +  
(\chi p y \theta)^2 }\, . \label{frel} 
\end{eqnarray} 
 
Equations (\ref{Qtot}) --  (\ref{frel}) determine 
frequency and temperature dependences of the ultrasound attenuation. 
We see that $W$ depends upon two tunable parameters, 
$\Omega$ and $\theta$. Thus the investigation of frequency and temperature 
dependences of  $Q_{\text{res}}$ provides a possibility to extract 
$E^*$ and $\chi$, see below Eqs.~(\ref{frela}), (\ref{fresa}), 
without an independent measurement of parameter $a$. 
 
First, we  analyze  the relaxation attenuation. Since in 
Eq.~(\ref{frel}) only $y \lesssim 1$ and $p < 1$  are important, 
to make an estimate one can  expand $\tanh y$  and neglect $p$ in 
$\sqrt{1-p}$. The result can be conveniently approximated as 
\begin{equation} 
F_{\text{rel}} \approx (\chi/\tilde{\Omega})\,  \min\{\theta,1\} \, ,
\quad \tilde{\Omega} = \max \{\Omega, \chi \theta\}\, . 
\label{frela}
\end{equation}
Similar 
approximation of Eq.~(\ref{fres})  yields 
\begin{equation} 
F_{\text{res}}= \left\{  
\matrix{ 
-\chi \ln \tilde{\Omega} + \Omega/\max\{\Omega, \theta\}, & \tilde{\Omega} 
  \ll 1 \cr  
\chi /\tilde{\Omega}^{2}, &\tilde{\Omega}  \gg 1  
} \right. \, . 
\label{fresa} 
\end{equation} 
As one can see, at high frequencies and temperatures  
the relaxation mechanism dominates, whereas at lowest temperatures 
the resonant mechanism is more important. It should be noted that the 
two contributions to $W$ can be effectively separated in experimental 
studies of nonlinear attenuation. Indeed, $W_{\text{res}}$ is 
well known~\cite{Hunklinger,Black}, to be suppressed  by relatively 
weak ultrasound, so  
that only the relaxation contribution remains.   
According to Eqs.~(\ref{frela}), the frequency and 
temperature dependences of the attenuation differ  qualitatively for 
$\hbar \omega \ge \chi T$ and $\hbar \omega \le \chi T$
and for $T < E^*$ and $T >E^*$. This should enable 
to determine the parameters $\chi$ and $E^*$. 
 
Let us compare the estimated value of $W$ with the acoustic 
attenuation in conventional metallic glasses $W_g$ assuming that the 
parameters $a$ and $\chi$ are the same. It turns out that  the 
resonant contributions are related as TLS densities of states at the 
energy  $\hbar \omega$. We have already concluded that the 
explanation~\cite{imry2} of the experiments~\cite{MW} 
requires the TLS density of states $n_T/E^*$ to exceed its typical 
value for  metallic glasses $P_g$ by {\em one-two orders of
magnitude}.
 As in metallic glasses the acoustic attenuation has been successfully 
measured, one should expect rather large effect. Comparing the 
relaxation contributions one finds $W_{\text{rel}}/W_g \approx 
(n_T/P_g \max\{E^*,T\}) \simeq 10 \div 10^2 {\rm min} (1,E^*/T)$.     
 
According to several publications, see Refs.~\onlinecite{CZ}
and~\onlinecite{2CK}  for a review, in real metals interaction of 
TLS with electrons may cause two-channel Kondo (2CK) effect with 
rather high (1-10 K) Kondo temperature, $T_K$.  For $T<T_K$, the 
crossection for inelastic scattering reaches the unitary limit, 
$\chi =1$ in Eq.~(\ref{tf}). It was suggested in
Ref.~\onlinecite{zawadowski1} that  
the experimentally observed temperature independent dephasing rate is 
a manifestation of the 2CK effect. We are  convinced, that possible 
values of $T_K$ in real metals are several orders of magnitude lower 
than the estimates of Refs.~\onlinecite{ZZ} and we presented a detailed 
discussion in a separate paper~\cite{unpublished}. 
 
Nevertheless, in a context of a  phenomenological speculation it
makes sense to examine the assumptions behind the explanation of
Ref.~\onlinecite{zawadowski1} and to discuss whether those assumptions
manifest themselves in other observable quantities such as the
specific heat and sound attenuation.
 
Let us adopt the simplistic model of Ref.~\onlinecite{zawadowski1} and
characterize  
the TLS's by randomly distributed Kondo temperatures $T_K$ 
and the level splitting $\Delta$. The distribution function 
${\cal P}$  for these quantities can be written as 
\begin{equation} 
{\cal P}(T_K, \Delta ) 
=(n_T/ \Delta^*T_K^*)\,  
P ( T_K/T_K^*, \Delta/\Delta^*)  
\label{P} 
\end{equation} 
where 
$T_K^*$ and $\Delta^*$ are some cut-off values for the Kondo
temperature and the splitting, respectively. 
The dimensionless function $P(x,y)$ is 
assumed to decay rapidly enough at $x \gg 1$ or $y \gg 1$ and to be 
normalized, $\int_0^\infty dx\int_0^\infty dy {P}(x,y) = 1$. 
In particular, we neglect the effects of the 
anisotropy in the coupling constants of TLSs with electrons. This 
simplification does not change  final results qualitatively. 
 
 Each TLS is characterized by its inelastic crossection, 
\begin{equation} 
\sigma_{\text{in}}^{(i)} = (4\pi/k_F^2)\,  {\tilde \chi}
\left(T_K^{(i)}/T,\Delta^{(i)}/T\right) \, ,
\label{i1} 
\end{equation} 
where superscript $(i)$ labels a particular TLS, and  
${\tilde \chi} \left(x,y\right)$ is a dimensionless function with the
following  
asymptotic behavior (we omitted numerical prefactors): 
\begin{eqnarray} 
&&{\tilde \chi}\left(x,y\right) = 
\left\{ 
\matrix{
1, & x \gg 1, \ y^2 \ll x; \cr
x^2/y^4, & x \gg 1, \ x \ll y^2 \ll x^2;\cr
L, &\text{otherwise};
}
\right.
\label{i3} \\
&&L\equiv \left[\ln \max\{1/x,y/x\}\, \max \left(1,y\,  \ln
\max\{1/x,y/x\} \right)\right]^{-2}\, .
\nonumber 
\end{eqnarray} 
Combining Eqs.~(\ref{i1}) and (\ref{P}), one obtains for the 
 dephasing time $\tau_\varphi$ the expression
\begin{equation} 
\frac{\tau_0}{\tau_\varphi}=  
\int dx dy \, P(x,y)\,  {\tilde \chi}\left(x\frac{T_K^*}{T},  
y\frac{\Delta^*}{T}  \right) 
\label{i4} 
\end{equation} 
where $1/\tau_0=4\pi n_Tv_F/k_F^2$ is the maximal rate of the electron 
scattering off a TLS. Using 
asymptotic expressions (\ref{i3}) in Eq.~(\ref{i4}) we find 
\begin{equation} 
\frac{\tau_0}{\tau_\varphi} = \left\{
\matrix{
\frac{\min\{1, T/\Delta^*\}} {\ln^2 \max \{T/T_K^*,\Delta^*/T_K^*\}}, &
T \gg T_K^*;\cr
\min \left\{1,\left(TT_K^*\right)^{1/2}/\Delta^*\right\}, & T \ll
T_K^* .
} \right.\label{i5}
\end{equation}
 According to Eq.~(\ref{i5}), the saturation in temperature dependence 
of $\tau_\varphi$ is possible {\em only} if $\Delta^* < T_K^*$. 
The corresponding temperature region is $(\Delta^*/T_K^*)^2 < T/T_K^*
< 1$. Temperature saturation of $\tau_\varphi$ was observed in
Ref.~\onlinecite{MW}   
in the interval 40 mK$<$T$<$1 K. This  indicates that the 
distribution of the level splittings required for the phenomenology of 
Ref.~\onlinecite{zawadowski1} is {\em not typical for metallic
glasses} but rather  
narrow with the upper boundary $\Delta^* \simeq 200$ mK. Therefore, 
the assumption about unusually narrow level splitting distribution is 
intrinsic both for Refs.~\onlinecite{imry2} and
\onlinecite{zawadowski1}. In this respect,  
the difference in physical assumptions in these references appears 
quantitative rather than qualitative --
both interpretations of the data\cite{MW} imply that  TLS's in the 
gold wires are essentially different from those in metallic glasses.


Accepting model \cite{zawadowski1},
let us now discuss the specific heat and acoustic absorption.
The specific heat  for
 a particular TLS $C^{(i)} \simeq (T/T_K^{(i)})\ln T_K^{(i)}/T$ at 
$(\Delta^{(i)})^2/T_K^{(i)} \ll T \ll T_K^{(i)}$. It yields the specific 
similar to Eq.~(\ref{cc}) with $\chi={\cal L} =1$ and $E^{*} \simeq
T_K*$. Therefore, the  TLS can produce the correction to the specific heat 
of the order of 2\%, which is difficult to detect. 
 
Sound attenuation of the TLS in Kondo regime happens mostly due to the 
relaxation mechanism. The relaxation rate (\ref{Gamma}) for the individual 
impurity $(i)$ in the regime $\min \left\{\Delta^{(i)}, 
(\Delta^{(i)})^2/T_K^{(i)}\right\} \ll T$ can be estimated as 
$\Gamma^{(i)} = \hbar^{-1}\max\left\{T/\ln^2(T/T_K^{(i)}), T_K^{(i)}\right\}$.
Using such rate 
and $p=0$ in Eq.~(\ref{Q2}),  
and averaging the result with the help of Eq.~(\ref{P}), we obtain 
for the most realistic range of frequencies $\hbar\omega \ll T$ 
\begin{equation} 
W \approx\frac{ n_T a^2\, \omega}{T} \frac{\hbar \omega}{T_K^*} 
\left\{ 
\matrix{
1, & T < T_K^*; \cr 
\ln^{-2}T/T_K^*, & T > T_K^* . 
} 
\right. 
\label{i7} 
\end{equation} 
This value is smaller than the corresponding result in glasses by a  
factor $\hbar\omega/T_K^*$, which once again makes it difficult to observe. 

 In conclusion, we demonstrated that if the concentration of TLS 
is sufficient to cause the observed dephasing rate, one also 
should expect quite substantial and specific effects in the 
specific heat $C$ and ultrasound attenuation  
$W$.  In both cases TLS 
contribution should dominate the electronic one at low enough 
temperatures.  
 These effects persist provided that $T \gg T_K^*$, where $T_{K}^*$ is
the mean Kondo temperature.  In the opposite limit,  
which corresponds to the developed 2 channel Kondo effect, TLS 
effects on $C$ and $W$ seem to be relatively small.  We do not 
think, however, that the limit $T \ll \bar T_{K}^*$ is possible 
to realize at the experimentally accessible temperatures,  see
Ref.~\onlinecite{unpublished} for the detailed discussion.  
For this reason we believe that the absence of the TLS 
contributions to $C$ and $W$ would mean their irrelevance for the 
dephasing rather than a realization of the 2 channel Kondo scenario. 
 

\widetext 
\end{document}